\documentclass[%
 superscriptaddress,reprint,
 amsmath,amssymb,
 aps,
]{revtex4-2}

\usepackage{xcolor}
\usepackage{soul}
\usepackage{graphicx}
\usepackage{dcolumn}
\usepackage{bm}

\newcommand{\appropto}{\mathrel{\vcenter{
  \offinterlineskip\halign{\hfil$##$\cr
    \propto\cr\noalign{\kern2pt}\sim\cr\noalign{\kern-2pt}}}}}

\newcommand{\bra}[1]{\langle #1 |}
\newcommand{\ket}[1]{| #1 \rangle}

\newcommand{\Schrodinger}{Schr\"{o}dinger }

\begin{document}

\preprint{APS/123-QED}

\title{A Real-Space Perspective on Dephasing in Solid-State High Harmonic Generation}
\author{Graham G. Brown}
\email{brown@mbi-berlin.de}
\affiliation{Max Born Institute, Max-Born-Stra\ss e 2A, 12489, Berlin, Germany}
\author{\'{A}lvaro Jim\'{e}nez-Gal\'{a}n}
\affiliation{Max Born Institute, Max-Born-Stra\ss e 2A, 12489, Berlin, Germany}
\affiliation{Joint Attosecond Science Laboratory, National Research Council of Canada and University of Ottawa, Ottawa, Canada}
\author{Rui E. F. Silva}
\affiliation{Max Born Institute, Max-Born-Stra\ss e 2A, 12489, Berlin, Germany}
\affiliation{ICMM, Centro Superior de Investigaciones Científicas, Madrid, Spain}
\author{Misha Ivanov}%
\affiliation{Max Born Institute, Max-Born-Stra\ss e 2A, 12489, Berlin, Germany}
\affiliation{Department of Physics, Humboldt University, Newtonstra\ss e 15, 12489 Berlin, Germany}
\affiliation{Blackett Laboratory, Imperial College London, London SW7 2AZ, United Kingdom}

\date{\today}

\begin{abstract}
We develop and demonstrate a fully real-space perspective on HHG in crystals. Due to Wannier-Stark localization induced on sub-cycle timescales in the presence of a strong field, real-space descriptions are natural for strongly driven solids.  Our approach allows us to address the origin of the extremely short dephasing times, which appear necessary for agreement between experimental HHG measurements and theoretical calculations generally performed in reciprocal space. We develop a physically transparent model of real-space dephasing which relates its rate to the distance between different sites in a laser-driven lattice. Our approach leads to well-structured high harmonic spectra at the microscopic level, reproduces results of macroscopic propagation, and demonstrates that the requirement for ultrafast dephasing times stems from the need for suppressing recombination events with large electron-hole separations during radiative recombination.
\end{abstract}

\maketitle


Since its discovery nearly a decade ago \cite{firstHHGSolid,Schubert2014}, high harmonic generation (HHG) in solids has led to major advances in the study of electron dynamics in solids. In particular, as a spectroscopic tool, HHG has been used to study band structure (both field-free \cite{allOpticalBandStructure} and laser-driven \cite{Uzan-Narovlansky2022}), density of states \cite{Uzan2020,PhysRevLett.118.087403}, electron-hole dynamics \cite{EH1,EH2}, multi-electron dynamics \cite{hhgManyBody,PhysRevA.105.053118}, the effects of Berry curvature \cite{HHGBC1,HHGBC2}, high-$T_c$ superconductivity \cite{https://doi.org/10.48550/arxiv.2201.09515}, coherent lattice dynamics \cite{PhysRevB.106.064303}, topological phase transitions \cite{HHGTPT}, and topological edge states \cite{Baykusheva2021,PhysRevLett.120.177401}.

In this context, a clear understanding of the microscopic and macroscopic components of the observed solid-state HHG is critical. At the microscopic level, the physical picture involves the complementary mechanisms of ($i$) intraband (e.g. \cite{firstHHGSolid,Goulielmakis2022}) and injection current driven emission \cite{Jurgens2020} and ($ii$) radiative electron-hole recombination, which extends the three-step recollision HHG model \cite{PhysRevLett.71.1994,PhysRevA.49.2117,Kulander1992} to solids as first proposed in \cite{vampa1}: following laser induced injection of a valence electron into the conduction band,  the electron and hole are accelerated by the field, until the electron-hole pair recombines to emit a photon. In contrast to atoms, in solids the relative distance between the recombining electron and hole need not be zero \cite{WQC}. These imperfect recollisions \cite{imperfectRecollisions,wannierHHGLewenstein} take advantage of phase coherence between different sites of the crystal and the delocalized nature of the Bloch states.

In this microscopic picture, a crucial issue remains outstanding: theoretically simulated HHG spectra show agreement with experiment only when extremely fast dephasing times $T_2 \sim 2$ fs are used \cite{vampa2}, in stark contrast to measurements \cite{dephasingTimes1,dephasingTimes2,dephasingTimes3} which show dephasing times on the scale of tens to hundreds of femtoseconds. Important steps towards solving this mystery have been made in \cite{PhysRevA.97.011401,PhysRevLett.125.083901,Abadie:18} by incorporating macroscopic propagation effects when simulating HHG. In particular, Refs. \cite{PhysRevA.97.011401,PhysRevLett.125.083901} argued that longitudinal propagation cleans up the harmonic spectra without assuming $\sim 1$ fs dephasing times. However, this so-called \emph{propagation-induced decoherence} \cite{PhysRevLett.125.083901} requires large propagation lengths of $\sim 10$ \textmu m, while many HHG experiments are done in very thin samples, including atomically thin monolayers \cite{HHGBC2}, still demonstrating highly regular HHG spectra. 

Here we show that the assumption of extremely rapid dephasing becomes unnecessary when a real-space description of solid-state HHG using the Wannier basis \cite{wannierOG,kohnWannier,RevModPhys.84.1419}  is used, rather than the conventional reciprocal-space description. A real-space description is natural for strongly driven solids, where Wannier-Stark localization is induced on the sub-cycle timescale \cite{PhysRevB.90.085313}. We show that the extreme complexity of the microscopic harmonic spectra is associated, first and foremost, with recombination events with large electron-hole separations \cite{WQC,imperfectRecollisions,wannierHHGLewenstein}. In the presence of a strong field, large electron-hole separations lead to large polarization-induced intensity-dependent phase shifts. These lead to the rapid divergence of the associated contributions to the harmonic emission, which disappear from the far-field signal upon spatial filtering \cite{Abadie:18}. Together with modest dephasing times ($T_2 \gtrsim 10$ fs), suppressing the contributions to HHG spectra from transitions involving large electron-hole separations leads to well-structured harmonic spectra already at the microscopic level, which faithfully reproduce the far-field signals as shown below. 
%
%

Let us briefly introduce our formalism, which uses maximally localized Wannier functions \cite{wannierOG,kohnWannier}. Let $\hat{\rho}(t)$ denote the time-dependent electronic density matrix and $\hat{H}(t)$ represent the full time-dependent electronic Hamiltonian including any external fields. The time-evolution of the density matrix elements $\rho_{m, m'}^{\alpha}(t)$ obeys the following master equation with a pure dephasing term \cite{masterEquation}:

\begin{equation}
	\frac{\partial \rho_{m, m'}^{\alpha}}{\partial t} = - i \left[ \hat{H} (t) , \hat{\rho} \right] \vphantom{x}_{m, m'}^{\alpha} - w_{m, m'}^{\alpha} \rho_{m, m'}^{\alpha},
	\label{eq:masterEQN}
\end{equation}

\noindent 
where $m$ and $m'$ denote the band indices, $\alpha$ denotes the crystal momentum $k$ (lattice site positions $R, R'$) when the Bloch (Wannier) basis is used, the interaction with the external field is described in the velocity gauge, and $w_{m, m'}^{\alpha}$ introduces decoherence into the system. 

We begin by describing the calculation in the Bloch basis. We do this for two reasons: ($1$) our Wannier model is constructed as a basis transformation of the Bloch basis model, and ($2$) we will compare our implementation of spatial dephasing in the Wannier basis with dephasing typically employed in the Bloch basis. We consider a one-dimensional periodic field-free Hamiltonian $\hat{H}_0$ whose eigenstates $\ket{\psi_{m, k}}$ can be described using the Bloch theorem such that

\begin{equation}
	\hat{H}_0 \ket{\psi_{m, k}} = \epsilon_m (k) \ket{\psi_{m, k}},
\end{equation}

\noindent 
where $k \in \left[ - \pi / a_0, \pi / a_0 \right]$ denotes the crystal momentum, $a_0$ is the lattice constant, $m$ denotes the band index, and $\epsilon_m (k)$ is the energy of the state in band $m$ with crystal momentum $k$. In this basis, the field-free Hamiltonian and momentum operators are expressed as follows (superscript $B$ denotes the Bloch basis):

\begin{align}
	\hat{H}_0^{(B)} &= \sum_{m} \sum_{k} \epsilon_m (k) \ket{\psi_{m, k}} \bra{\psi_{m, k}}, \\
	\hat{p}^{(B)} &= \sum_{m, m'} \sum_{k} p_{m, m'}^{k} \ket{\psi_{m, k}} \bra{\psi_{m', k}} ,
\end{align}

\noindent 
where $p_{m, m'}^{k} = \bra{\psi_{m, k}} \hat{p} \ket{\psi_{m', k}}$. 

In the Bloch basis, the decoherence term $w_{m, m'}^{k}$ is typically defined as a uniform decay of all coherences between the valence and conduction band with a characteristic decay time known as the dephasing time $T_2$ \cite{vampa2}:

\begin{equation}
	w_{m, m'}^{k} = \frac{1 - \delta_{m, m'}}{T_2}  , 
	\label{eq:dephasingBloch}
\end{equation}

\noindent
where $\delta_{m, m'}$ is the Kronecker delta function. 

We now describe our model in the Wannier basis. The Wannier orbital in band $m$ at lattice site $R = n a_0$ (integer $n$) is calculated from the Bloch basis as follows \cite{kohnWannier}:

\begin{equation}
	\ket{\phi_{m, R}} = \frac{1}{\sqrt{N}} \sum_k e^{- i k R} \ket{\psi_{m, k}},
	\label{eq:wannierBlochTransform}
\end{equation}

\noindent 
where $N$ is the number of lattice sites in the system and we have followed the procedure outlined in \cite{kohnWannier} in order to obtain maximally localized Wannier functions. From Eq. (\ref{eq:wannierBlochTransform}), the field-free Hamiltonian and momentum operators in the Wannier basis are given as follows (superscript $W$ denotes the Wannier basis):

\begin{align}
	\hat{H}_0^{(W)} &= \sum_{m,k} \sum_{R, R'} e^{i k \Delta R} \epsilon_m (k) \ket{\phi_{m, R}} \bra{\phi_{m, R}}, \\
	\hat{p}^{(W)} &= \sum_{m, m',k} \sum_{R, R'} p_{m, m'}^{k} e^{i k \Delta R}	 \ket{\phi_{m, R}} \bra{\phi_{m', R'}},
\end{align}

\noindent 
where $\Delta R = R - R'$. 

\begin{figure}[t]
	\includegraphics{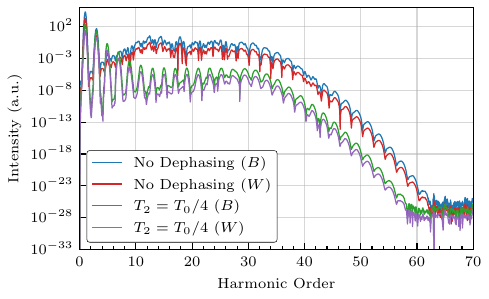}
	\caption{The high harmonic spectra calculated without dephasing and with a dephasing time of one quarter laser cycle, $T_0 / 4 = 2.7$ fs, using Eq. (\ref{eq:dephasingBloch}) from the Bloch (blue and green, respectively) and the Wannier (red and purple, respectively) bases. The spectra from the Wannier basis are shifted vertically $\times 10^{-1}$ for clarity and  the driving field is an eight-cycle Gaussian pulse with peak intensity $I_0 = 3.1 \times 10^{11}$ W/cm$^{2}$ and wavelength $\lambda = 3.2$ \textmu m.}
	\label{fig:wannierBlochHHGComparisonWithoutDecoherence}
\end{figure}

Fig. \ref{fig:wannierBlochHHGComparisonWithoutDecoherence} shows the equivalence of the HHG spectra calculated from the Bloch and Wannier bases with and without dephasing implemented using Eq. (\ref{eq:dephasingBloch}). For all calculations, we consider a two-band one-dimensional model with lattice constant $a_0 = 4.2$ \AA (8 a.u.) and a periodic Mathieu-type potential with a strength of 10.1 eV ($0.37$ a.u.) as in \cite{hhgFromBEiS}. We use a one-dimensional model to unambiguously isolate the effects of spatial dephasing from effects related to crystal structure. We simulate the interaction of our system with an eight-cycle Gaussian pulse driving field with a peak intensity $3.1 \times 10^{11}$ W/cm$^{2}$ ($F_0 = 0.003$ a.u.) and wavelength $3.2$ \textmu m. For long wavelength driving lasers, it is expected that interband emission is the dominant HHG mechanism \cite{vampa3}. The spectra calculated without dephasing from the Bloch and Wannier bases are shown by the blue and red lines, respectively. The spectra calculated with a dephasing time of one quarter optical cycle $T_0 / 4 = 2.7$ fs using Eq. (\ref{eq:dephasingBloch}) from the Bloch and Wannier bases are shown by the green and purple lines, respectively. Both spectra calculated using the Wannier basis are shifted $\times 10^{-1}$ vertically for clarity. As expected, the two spectra are indistinguishable. 

We now introduce the spatially resolved matrix $w_{m, m'}^{R, R'}$ used in the Wannier basis. First, however, we briefly discuss methods for obtaining clear harmonic spectra from single-atom simulations of gas-phase HHG to motivate our model \cite{doi:10.1063/1.1517042,coherentSum}. For gas-phase HHG, spectra calculated by solving the time-dependent \Schrodinger equation (TDSE) do not exhibit clear harmonic peaks unless either a complex-valued absorbing potential is used or the single-atom simulations are supplemented with simulations of macroscopic propagation. Both methods suppress the contribution of long trajectories and high-order returns of the continuum electron wavepacket to its parent ion in the harmonic spectra. 

The emergence of clear harmonic spectra upon propagation results from the coherent summation of the dipole emission from single-atom simulations calculated with a range of driving field intensities \cite{coherentSum}. Over this range of intensities, the phase of emission associated with high-order returns varies by more than $2 \pi$. When the coherent summation is performed, the large variation in phase from high-order returns leads to their destructive interference, resulting in a clear harmonic structure. 

Returning to solid-state HHG described in the Wannier basis, we now make an analogous argument where, instead of high-order returns, we focus on imperfect recollisions with large electron-hole separations specific for solids.

Unlike atomic HHG, HHG in periodic systems involves emission from an infinite number of sources corresponding to each possible electron-hole separation at the time of recombination. To see this, one can use the Wannier basis to express the current between bands $m$ and $m'$ as a summation over the possible electron-hole separations $\Delta n a_0$ for integer $\Delta n$ at the time of recombination:

\begin{equation}
	\begin{split}
		j_{m, m'}(t) &\propto a_{m}^*(t) a_{m'}(t) \sum_{\Delta n, k} e^{i (k + A(t)) \Delta n a_0} p_{m, m'}^{k + A(t)},
	\end{split}
	\label{eq:j_mmp}
\end{equation}

\noindent 
plus the complex conjugate, where $a_{m}(t)$ is the amplitude of each Wannier orbital in band $m$. The current in Eq. (\ref{eq:j_mmp}) consists of a sum over all the possible electron-hole separations at the time of recombination, each with a phase difference of $\Delta \Phi(\Delta n) = (k + A(t)) \Delta n a_0$.

The characteristic scale of the field-dependent contribution to the phase can be estimated as $\Delta \Phi (\Delta n) \sim A_0 a_0 \Delta n / 2$. Destructive interference will inevitably suppress contributions to experimentally measured spectra from imperfect recollisions with separations greater than $\Delta n_2$ lattice sites whenever $\Delta \Phi (\Delta n_2) > 2 \pi$, just as seen for atomic media \cite{coherentSum,Gaarde_2008}. We now show that suppressing coherences in $\hat{\rho}^{(W)}$ for which the electron-hole separation is greater than $\Delta n_2$ while maintaining long dephasing times for coherences with smaller separations will result in clear harmonic spectra. Accordingly, we implement the decoherence matrix $\hat{w}^{(W)}$ in real space as

\begin{equation}
	\begin{split}
		\hat{w}^{R, R'}_{m, m'} &= \left(1 - \delta_{m, m'}\right) \left( \frac{1}{T_2} + \Gamma\left(\Delta n\right) \right),
	\end{split}
	\label{eq:wannierW}
\end{equation}

\noindent 
where $T_2$ is the conventional dephasing time and $\Gamma(\Delta n)$ is zero for lattice-site separations below $\Delta n_2$ and increases rapidly when $| \Delta n | \ge \Delta n_2$. Below, we use the following:

\begin{equation}
	\Gamma \left( \Delta n \right) = \frac{1}{T_0} \begin{cases} \left(\frac{| \Delta n | - \Delta n_2}{\sigma} \right)^2 & | \Delta n | \ge \Delta n_2 \\ 0 & \mbox{otherwise}
	 \end{cases} ,
	\label{eq:wannierWF}
\end{equation}

\noindent 
where $\Delta n_2$ is the dephasing boundary, $T_0$ is the optical cycle of the driving field, and $\sigma = 6$, resulting in an effective dephasing time of one quarter optical cycle ($2.7$ fs) when $|\Delta n| = \Delta n_2 + 6$. o	These parameters are chosen to obtain agreement between microscopic and far-field macroscopic HHG spectra (discussed below). For our simulations, $A_0 = 0.21$ a.u. and $a_0 = 8$ a.u., resulting in a predicted dephasing boundary of $\Delta n_2 = 8$ which satisfies $\Delta \Phi(\Delta n_2) > 2 \pi$. The dephasing time $T_2$ in Eq. (\ref{eq:wannierW}) is set to $1000$ fs, which results in negligible uniform dephasing of excitations. 

\begin{figure}[t]
	\includegraphics{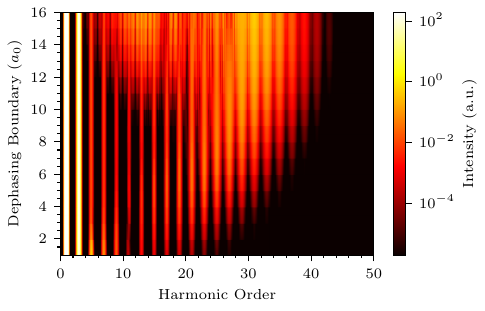}
	\caption{The HHG spectra calculated from the Wannier basis with dephasing implemented using Eqs. (\ref{eq:wannierW}) and (\ref{eq:wannierWF}) for dephasing boundaries $\Delta n_2 \in \left[ 1 , 16 \right]$. The driving field is the same as used to generate the spectra in Fig. \ref{fig:wannierBlochHHGComparisonWithoutDecoherence}.}
	\label{fig:variationLR}
\end{figure}

The HHG spectra calculated from the Wannier basis using this model of decoherence are shown in Fig. \ref{fig:variationLR} for $\Delta n_2 \in [ 1, 16]$. The maximum generated photon energy increases with the dephasing boundary until $\Delta n_2 = 11$ due to the increasing importance of imperfect recollisions with large electron-hole separations, which emit higher-energy photons due to the large polarization energy induced by the external field. For $\Delta n_2 \le 10$, the spectra exhibit distinct harmonic peaks. Based on our simulation parameters, electron-hole separations larger than eight lattice sites will exhibit polarization-induced phase shifts exceeding $2 \pi$. As the dephasing boundary extends beyond eight lattice sites, chaotic structure rapidly increases and the harmonic structure of the plateau is completely lost for dephasing boundaries $\Delta n_2 \ge 14$. These results show that clear harmonic spectra can be obtained while maintaining near full coherence for lattice site separations below $\Delta n_2$ when coherences between distantly separated lattice sites are suppressed. 

\begin{figure}
	\includegraphics{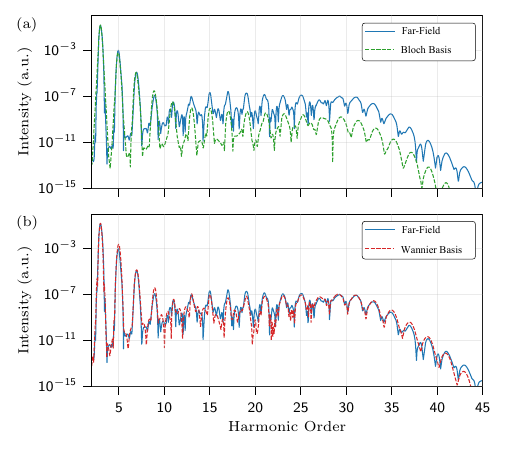}	
	\caption{Comparison of the HHG spectrum obtained after far-field propagation (blue) with HHG spectra calculated with (a) a uniform dephasing time $T_2 = 2.7$ fs (dashed green) in the Bloch basis and (b) the spatially-dependent dephasing mechanism given by Eqs. (\ref{eq:wannierW}) and (\ref{eq:wannierWF}) with $\Delta n_2 = 8$ and $\sigma = 6$ (dashed red) in the Wannier basis.}
	\label{fig:farFieldComparison}
\end{figure}

We now compare HHG spectra calculated with both considered dephasing mechanisms with spectra obtained after simulating far-field propagation. Fig. \ref{fig:farFieldComparison} shows the comparison of the HHG spectra calculated with (a) a uniform dephasing time of  $T_2 = 2.7$ fs (dashed green line) and (b) the spatially-dependent dephasing mechanism given by Eqs. (\ref{eq:wannierW}) and (\ref{eq:wannierWF}) (dashed red line) with the spectrum obtained after far-field propagation (solid blue line in both subfigures). The spatially-dependent dephasing mechanism is calculated with a dephasing boundary of $\Delta n_2 = 8$ and width of $\sigma = 6$ lattice sites. The far-field spectrum is obtained by simulating HHG with a dephasing time of $T_2 = 10$ fs across the beamfront of a radially symmetric Gaussian beam with a waist of 50 \textmu m at its focus, propagating the resultant HHG emission a distance of 1 m into the far-field \cite{Guizar-Sicairos:04}, and radially integrating the spectral beamfront. All spectra are normalized according to the integrated HHG emission above harmonic order 3.

From Fig. \ref{fig:farFieldComparison} (a), it is immediately apparent that the use of ultrafast dephasing times significantly underestimates HHG emission above the minimum band gap near harmonic order 11 when compared with the spectrum obtained after simulating far-field propagation. This can be understood by recognizing that ultrafast dephasing times $\sim 1$ fs are comparable to some recollision trajectory excursion times. The use of ultrafast dephasing times suppresses these trajectories and, therefore, their respective HHG emission which otherwise survives far-field propagation. In contrast, the spectrum calculated with a spatially-dependent dephasing mechanism shown in (b) agrees with the spectrum obtained after simulating far-field propagation for all harmonic orders. By maintaining nearly full coherence for lattice site separations below the dephasing boundary, only highly divergent HHG emission from distantly-separated coherences which is not observed experimentally is suppressed and there is no artificial suppression of trajectory dynamics within the dephasing boundary.

We emphasize that our approach to dephasing is not a phenomenological incorporation of microscopic dynamics beyond the independent-particle approximation. The decoherence introduced by $\hat{w}^{(W)}$ is instead motivated by the expectation that destructive interference will suppress emission from large electron-hole separations which exhibit large polarization-induced phase differences. In this regard, our approach is analogous to the use of complex absorbing potentials to mimic propagation effects \cite{doi:10.1063/1.1517042} and obtain clear harmonic spectra in gas-phase HHG. At the same time, our approach opens a natural avenue for introducing microscopic dephasing mechanisms associated with crystal imperfections, other electrons, or phonons, as these are naturally expressed via real-space free paths. When expressed in reciprocal space, space-dependent dephasing naturally leads to dephasing rates growing as crystal momenta move far away from the minimum band gap. 

In conclusion, we have demonstrated that the ultrafast dephasing times required for previous descriptions of HHG in solids \cite{vampa2} are entirely unnecessary. When described in the Wannier basis, which is natural in strong electric fields, the requirement for effective ultrafast dephasing times follows from the suppression of distantly separated coherences. Our results suggest that decoherence lengths are more appropriate to characterize strong-field processes in solids than decoherence times.

The spatially-dependent dephasing mechanism can be understood within the context of \cite{Abadie:18}, which first demonstrated that clear harmonic spectra can be obtained by filtering out highly divergent components of the HHG emission in the far-field. We show that the highly divergent HHG emission is related to emission from coherences between distantly separated lattice sites (e.g. \emph{imperfect recollisions}). We suppress those while maintaining near full coherence between lattice sites with small separations, thereby retaining the components of HHG emission which dominate experimentally measured spectra.

G. G. Brown acknowledges funding from the European Union’s Horizon 2020 research and innovation programme under grant agreement No. 899794 (Optologic). M. Ivanov  acknowledges funding from the SFB 1477 ``Light Matter Interaction at Interfaces" project number 441234705. \'{A}.J.G. acknowledges funding from the European Union’s Horizon 2020 research and innovation programme under the Marie Skłodowska-Curie grant agreement no. 101028938. \'{A}.J.G. acknowledges funding from the European Union’s Horizon 2020 research and innovation programme under the Marie Skłodowska-Curie grant agreement no. 101028938. R. E. F. Silva acknowledges support from the fellowship LCF/BQ/PR21/11840008 from ``La Caixa” Foundation (ID 100010434). We thank A. Marini, H. Gross, E. Goulielmakis, and A. Leitenstorfer for exceptionally useful comments.

\bibliography{apssamp}

\end{document}